\documentclass[aps,prd,reprint,superscriptaddress,floatfix]{revtex4-2}
\usepackage{amsmath,amssymb,bm}
\usepackage{graphicx,xcolor,booktabs}
\usepackage[hidelinks]{hyperref}
\usepackage{orcidlink}
\definecolor{revisionblue}{RGB}{0,0,0}
\newif\ifshowchanges
\showchangestrue
\ifdefined\CleanCopy\showchangesfalse\fi
\newenvironment{revision}{\begingroup\ifshowchanges\color{revisionblue}\fi}{\endgroup}
\DeclareRobustCommand{\rev}[1]{{\ifshowchanges\color{revisionblue}\fi#1}}
\newcommand{\revisioncolor}{\ifshowchanges\color{revisionblue}\fi}
\begin{document}
\title{Noncommutative Kazakov-Solodukhin Gravastar}
\author{A. T. N. Silva~\orcidlink{0009-0007-0067-600X}}
\email{andersont.nsilva@gmail.com}
\affiliation{Departamento de F\'{\i}sica, Universidade Federal de Campina Grande
Caixa Postal 10071, 58429-900 Campina Grande, Para\'{\i}ba, Brazil}
\author{M. A. Anacleto~\orcidlink{0000-0003-4625-7322}}
\email{anacleto@df.ufcg.edu.br}
\affiliation{Departamento de F\'{\i}sica, Universidade Federal de Campina Grande
Caixa Postal 10071, 58429-900 Campina Grande, Para\'{\i}ba, Brazil}
\affiliation{Unidade Acad\^emica de Matem\'atica, Universidade Federal de Campina Grande
\\
58429-900 Campina Grande, Para\'{\i}ba, Brazil}
\author{Ali \"Ovg\"un~\orcidlink{0000-0002-9889-342X}}
\email{ali.ovgun@emu.edu.tr}
\affiliation{Physics Department, Faculty of Arts and Sciences,
Eastern Mediterranean University,
Famagusta, 99628 North Cyprus via Mersin 10, T\"urkiye}
\author{L. Casarini~\orcidlink{0000-0002-0869-9405}}
\email{lcasarini@academico.ufs.br}
\affiliation{Departamento de F\'isica, Universidade Federal de Sergipe
49100-000 Aracaju, Sergipe, Brazil}

\begin{abstract}
\begin{revision}
We construct a thin-shell gravastar with a de Sitter interior and a
phenomenological exterior combining a Lorentzian noncommutative-inspired
mass distribution with the unexpanded Kazakov--Solodukhin deformation.
Within an effective Einstein description, we derive the surface stresses,
evaluate the shell energy conditions, and compute transparent-shell
geodesics using a single exterior metric. The associated horizon-bearing
family satisfies a radial-pressure thermodynamic identity with area
entropy. For the gravastar, the first three penetrating visibility
boundaries move inward with increasing KS scale, while the exterior
photon orbit and critical impact parameter move outward. A finite-observer
backlighting calculation reproduces the Schwarzschild--de Sitter
benchmark and tracks the merger of visibility boundaries as the shell
moves beyond the photon orbit. We also compute the redshift-weighted
intensity of a face-on, optically thin emitting layer. Additional disk
crossings after core transmission produce inner image structure whose
brightness depends on the radial emission profile. At leading
inverse-radius order, both deformation scales enter through one
coefficient, producing an exact geometric parameter degeneracy.
We show that the next inverse-radius term breaks this degeneracy:
the critical impact parameter and logarithmic strong-deflection
coefficient provide independent local information about the two scales.
Full-metric calculations for equal-leading-coefficient parameter pairs
resolve this splitting and quantify the accuracy of the higher-order
expansion.
\end{revision}
\end{abstract}
\maketitle
\section{\rev{Introduction}}
\begin{revision}
Compact objects with a regular interior provide a way to examine which
strong-field signatures depend on an event horizon and which follow from
the exterior gravitational field. Horizon-scale images
\cite{EventHorizonTelescope:2019dse,EventHorizonTelescope:2022wkp}
and gravitational-wave observations \cite{KAGRA:2021duu} motivate this
distinction. A gravastar replaces the central black-hole region by a
de Sitter core supported by an intervening matter layer
\cite{Mazur:2001fv,Mazur:2004fk}. In the thin-shell description, the
Israel junction conditions determine the surface density and pressure
from the geometries on either side of the shell
\cite{Israel,Visser:2003ge,MartinMoruno:2011rm}.

The shell properties, perturbative response and optical appearance probe
different aspects of this construction. Energy conditions constrain the
surface stresses and compactness \cite{Horvat:2007qa}; nonradial
oscillations and inspiral signals depend on the dynamical response
\cite{Pani:2009ss,Pani:2010em}; tidal deformability probes the response
to an external field \cite{Uchikata:2016qku}. For a transparent object,
light can cross the shell, turn inside the core and return to the
observer. Illumination by a background optical plane then produces
a central bright region and alternating bright and dark rings
\cite{Sakai:2014pga,Kubo:2016ada}. Accretion disks and orbiting hot spots
give complementary images with their own source dependence
\cite{Rosa:2024bqv}. For an optically thin emitting disk, the image
follows from the radii of successive disk crossings and their
redshift-weighted intensities \cite{GrallaHolzWald2019}.
Additional returning branches modify these images in horizonless
geometries \cite{PengGuoFengWormhole2021}.

Two prescriptions introduce short-distance scales into the exterior
geometry considered here. A noncommutative-inspired distribution replaces
a point source by an extended density, as in the Gaussian Schwarzschild
model \cite{Nicolini:2005vd}. We use a Lorentzian profile, whose integrated
mass approaches the ADM mass at spatial infinity
\cite{Anacleto:2019tdj}. Gravastars built from smeared sources have been
studied in four and three dimensions
\cite{Lobo:2010uc,BANERJEE2016,Ovgun:2017jzt}, including BTZ models
\cite{SilvaAnacletoCasarini2024,AnacletoSilvaCasarini2025}.
The noncommutative BTZ gravastar construction was
subsequently extended to include Gauss--Bonnet
corrections, with an analysis of the shell energy
conditions and thermodynamic properties
~\cite{AnacletoSilvaCasariniCarvalho2026}.

The optical appearance of the four-dimensional Lorentzian model was
examined in Ref.~\cite{Anacleto:2026hkb}.
Related investigations have addressed quasinormal modes,
absorption, scattering, and shadows in noncommutative
black-hole geometries
~\cite{Campos:2021sff,Anacleto:2022shk}.
Geodesic motion, gravitational lensing, and accretion-disk
images have also been examined in modified black-hole
spacetimes~\cite{AraujoFilho:2024mvz}.

The Kazakov--Solodukhin (KS) deformation originates in an effective
two-dimensional dilaton treatment of spherically symmetric gravity
\cite{KazakovSolodukhin1994}. Its additional length scale changes the
exterior photon orbits, scattering and quasinormal spectrum
\cite{KonoplyaKS2020}, as well as the lensing rings and images of
accretion disks \cite{PengGuoFeng2021,HuangEtAl2023KS}.
For the unexpanded KS metric, the Hawking temperature at fixed ADM mass
equals the Schwarzschild value, even though the horizon radius changes
\cite{BolokhovBronnikovKonoplya2025}. This distinction between fixed
mass and fixed horizon radius is relevant when comparing deformed
families of compact objects.
Thermodynamic corrections associated with
noncommutative geometry and the generalized
uncertainty principle have also been investigated
in Schwarzschild and BTZ geometries
~\cite{Anacleto:2020zfh,Anacleto:2020efy}.

We combine the Lorentzian mass profile and the unexpanded KS factor in
a phenomenological exterior, interpreted through the Einstein equations
with an effective anisotropic source. This choice supplies a common
framework for the bulk stress tensor, the shell junction conditions and
the area entropy of the associated horizon-bearing family. A de Sitter
core replaces the inner region, and the primary shell family lies in a
globally static exterior below the unstable photon orbit. Our primary numerical
predictions use the unexpanded metric, with truncated metrics providing
controlled comparisons. We reproduce the Schwarzschild--de Sitter
backlighting benchmark of Ref.~\cite{Sakai:2014pga}, retain a finite
observer distance, and follow the ring mergers when the shell radius
extends beyond the exterior photon orbit.
We then compute disk transfer functions and color intensity maps for
two explicit emission profiles. Comparing the same disk around the
gravastar and its associated black hole isolates the light received
after transmission through the core.

The principal question is whether the two deformation scales can be
separated by the optical observables. Their leading contributions enter
the metric through a single inverse-square coefficient. We show that
this produces an exact parameter degeneracy in the leading exterior
model, including its shell stresses and transparent-shell images. The
next inverse-radius term supplies independent information: the critical
impact parameter and the logarithmic strong-deflection coefficient
respond differently to it. We derive this response and resolve the
splitting numerically for two parameter pairs with the same leading
coefficient. The strong-deflection analysis uses the exterior formalism
of Ref.~\cite{Bozza2002NCaudit} and the established inner-image
coefficient for ultracompact objects \cite{Shaikh2019NCaudit}.

Section~\ref{sec1} defines the exterior and its effective source.
Section~\ref{sec:thermodynamics} derives the horizon thermodynamic
identities. Section~\ref{sec:structure} constructs the shell and tests
its energy conditions. Section~\ref{sec:null_geodesics} develops the
transparent-shell ray tracing. Section~\ref{sec:disk_images} constructs
the disk transfer functions and observed intensities, and
Sec.~\ref{sec:optical_identifiability} examines the higher-order
separation of the deformation scales. Section~\ref{Conc} gives the
conclusions.
\end{revision}

\section{\rev{Exterior geometry and effective source}}
\label{sec1}
\begin{revision}
We use signature $(-,+,+,+)$ and $G=c=1$. The exterior line element is
\begin{equation}
 ds_+^2=-F(r)dt_+^2+\frac{dr^2}{F(r)}+r^2d\Omega^2,
 \label{eq:ks_metric}
\end{equation}
where $r$ is the areal radius and
$d\Omega^2=d\vartheta^2+\sin^2\vartheta\,d\varphi^2$. The Lorentzian
density and its integrated mass are
\begin{align}
 \rho_\theta(r)&=\frac{M\sqrt\theta}
 {\pi^{3/2}(r^2+\pi\theta)^2},\label{MD1}\\
 \mathcal M_\theta(r)&=4\pi\int_0^r \rho_\theta(u)u^2du=M\mu(r),
 \label{eq:nc_mass}\\
 \mu(r)&=\frac2\pi\left[\arctan\frac r w-
                  \frac{rw}{r^2+w^2}\right],\qquad
 w=\sqrt{\pi\theta}.
 \label{eq:mu_definition}
\end{align}
Thus $\mu(\infty)=1$ and $M$ is the ADM mass. At $\theta=0$ we use
the continuous limit $\mu(r)=1$ for $r>0$.

Our exterior prescription is
\begin{equation}
 F(r)=\sqrt{1-\frac{\eta^2}{r^2}}-\frac{2M\mu(r)}r,
 \qquad r>\eta,
 \label{eq:full_lapse}
\end{equation}
with $M>0$, $\theta\geq0$ and $\eta\geq0$. The parameters
$\sqrt\theta$ and $\eta$ have dimensions of length. The limits
$\theta=0$, $\eta=0$ and $\theta=\eta=0$ give the unexpanded KS,
Lorentzian noncommutative-inspired and Schwarzschild exteriors,
respectively.

We adopt an Einstein--Hilbert gravitational sector with a minimally
coupled effective source and represent the deformation by the stress tensor
$T^{\rm eff}_{\mu\nu}=G_{\mu\nu}/(8\pi)$. Defining
\begin{equation}
 g(r)=\sqrt{1-\frac{\eta^2}{r^2}},\qquad
 m_{\rm eff}(r)=M\mu(r)+\frac r2[1-g(r)],
 \label{eq:effective_mass}
\end{equation}
so that $F=1-2m_{\rm eff}/r$, the source components are
\begin{align}
 \rho_{\rm eff}&=\frac{m_{\rm eff}'}{4\pi r^2}
  =\rho_\theta+\frac{1-g^{-1}}{8\pi r^2},\nonumber\\
 p_r&=-\rho_{\rm eff},\qquad
 p_t=-\frac{m_{\rm eff}''}{8\pi r}.
 \label{eq:effective_stress}
\end{align}
The KS term therefore contributes to the total anisotropic source in
addition to the Lorentzian density. This effective description fixes
the gravitational junction conditions and the entropy convention used
below. The gravastar replaces the region below its shell by a regular
de Sitter core.

For analytic comparisons, introduce
\begin{equation}
 s=\sqrt{\theta/\pi},\qquad e=\eta^2,\qquad
 q=8Ms-\frac e2.
 \label{eq:id_parameters}
\end{equation}
Expansion of Eq.~\eqref{eq:full_lapse} gives
\begin{align}
 F(r)&=1-\frac{2M}r+\frac q{r^2}
           +\frac D{r^4}+\frac E{r^6}+O(r^{-8}),
 \label{eq:id_expansion}\\
 D&=-\frac{16\pi^2Ms^3}{3}-\frac{e^2}{8},\qquad
 E=\frac{24\pi^4Ms^5}{5}-\frac{e^3}{16}.
 \label{eq:id_DE}
\end{align}
The respective dimensions of $q,D,E$ are $L^2,L^4,L^6$.
The expansion is controlled by $\pi s/r\ll1$ and $\sqrt e/r\ll1$.
Its leading form,
\begin{equation}
 F_q(r)=1-\frac{2M}r+\frac q{r^2},
 \label{FNC}
\end{equation}
contains the two microscopic scales only through $q$. We use
Eq.~\eqref{FNC} for explicit analytic limits and
Eq.~\eqref{eq:full_lapse} for the primary numerical predictions.
The higher-order analysis also evaluates truncated metrics to quantify
their accuracy against the full exterior.
\end{revision}

\section{\rev{Thermodynamics of the associated horizon family}}
\label{sec:thermodynamics}
\begin{revision}
We consider the horizon-bearing family associated with the same exterior
metric used in the gravastar construction. Throughout this section
$G=c=\hbar=k_B=1$, and the dimensional parameters $\theta$ and $\eta$
are held fixed as the ADM mass varies. We denote the outer horizon radius
by $R$. The functions $g(r)$ and $\mu(r)$ are those in
Eqs.~\eqref{eq:effective_mass} and~\eqref{eq:mu_definition}, so that
$F(r)=g(r)-2M\mu(r)/r$.
The physical branch has $R>\eta$, $F(R)=0$ and $F(r)>0$ for all $r>R$.

\subsection{\rev{Horizon mass and temperature}}

The horizon equation determines the mass as
\begin{equation}
 M(R)=\frac{R g(R)}{2\mu(R)}.
 \label{eq:thermo_mass_exact}
\end{equation}
The temperature follows from the surface gravity, with the radial
derivative of $F$ evaluated at fixed $M,\theta,\eta$:
\begin{align}
 T_H&=\frac1{4\pi}\left.\frac{\partial F}{\partial r}\right|_{r=R}
 \nonumber\\
 &=\frac1{4\pi}\left[g'(R)+\frac{g(R)}R
       -g(R)\frac{\mu'(R)}{\mu(R)}\right].
 \label{eq:thermo_temperature_exact}
\end{align}
The derivatives required in this expression are
\begin{equation}
 g'(r)=\frac{\eta^2}{r^3g(r)},\qquad
 \mu'(r)=\frac{4w r^2}{\pi(r^2+w^2)^2},\qquad
 w=\sqrt{\pi\theta}.
 \label{eq:thermo_profile_derivatives}
\end{equation}
Equations~\eqref{eq:thermo_mass_exact} and
\eqref{eq:thermo_temperature_exact} retain the full Lorentzian profile
and the unexpanded KS factor.

For comparison with the large-radius approximation, let
$k=4\sqrt{\theta/\pi}$. The inverse-square lapse gives
\begin{align}
 M_{\rm tr}(R)&=\frac{R^2-\eta^2/2}{2(R-k)},\nonumber\\
 T_{\rm tr}(R)&=\frac{R^2-2kR+\eta^2/2}{4\pi R^2(R-k)}.
 \label{eq:thermo_truncated_consistent}
\end{align}
To first order in $k/R$ and $\eta^2/R^2$,
\begin{equation}
 T_H(R)=\frac1{4\pi R}
 \left[1-\frac{k}{R}+\frac{\eta^2}{2R^2}
 +O\!\left(\frac{k^2}{R^2},\frac{k\eta^2}{R^3},
                         \frac{\eta^4}{R^4}\right)\right].
 \label{eq:thermo_temperature_radius_series}
\end{equation}
At fixed ADM mass the horizon displacement
$R=2M-k+\eta^2/(4M)+\cdots$ cancels the individual first-order
corrections in the temperature. The pure KS limit provides an exact check:
\begin{equation}
 R=\sqrt{4M^2+\eta^2},\qquad
 T_H=\frac1{8\pi M}\quad (\theta=0).
 \label{eq:thermo_KS_limit}
\end{equation}
This fixed-mass result agrees with the KS analysis of
Ref.~\cite{BolokhovBronnikovKonoplya2025}.

\subsection{\rev{Area entropy and heat response}}

With an Einstein--Hilbert gravitational sector and an effective source
without explicit curvature couplings, the Wald entropy reduces to the
area law~\cite{Wald1993},
\begin{equation}
 S_A=\frac{A_H}{4}=\pi R^2.
 \label{eq:thermo_area_entropy}
\end{equation}
The radial Einstein equation gives a local thermodynamic
identity~\cite{Padmanabhan2002}.
With $E_H=R/2$, $V_H=4\pi R^3/3$ and the radial pressure at the horizon
\begin{equation}
 P_H=\frac{R\,\partial_rF(R)-1}{8\pi R^2},
 \label{eq:thermo_horizon_pressure}
\end{equation}
one obtains
\begin{equation}
 dE_H=T_H\,dS_A-P_H\,dV_H.
 \label{eq:thermo_horizon_first_law}
\end{equation}
Differentiating the horizon equation along the fixed-$\theta,\eta$
family also yields
\begin{equation}
 \mu(R)\frac{dM}{dR}=2\pi R T_H,
 \qquad \mu(R)\,dM=T_H\,dS_A.
 \label{eq:thermo_mass_profile_identity}
\end{equation}
The factor $\mu(R)$ tracks the variation of the distributed source as
the ADM mass changes. Source-dependent factors of this type arise when
thermodynamic variations are restricted to prescribed metric families
\cite{MaZhao2014,ToshmatovEtAl2026}. Here the entropy and pressure work
are fixed by Eqs.~\eqref{eq:thermo_area_entropy}--
\eqref{eq:thermo_horizon_first_law}, while
Eq.~\eqref{eq:thermo_mass_profile_identity} specifies their relation to
the ADM-mass parametrization. The deformation dependence of $S_A$ at fixed
mass enters through the solution $R(M,\theta,\eta)$ of the horizon equation.

We characterize the fixed-deformation horizon family by the entropy heat
response and the ADM-mass slope,
\begin{equation}
 C_A\equiv T_H\frac{dS_A}{dT_H}
       =\frac{2\pi R T_H}{T_H'(R)},\qquad
 C_M\equiv\frac{dM}{dT_H}=\frac{C_A}{\mu(R)}.
 \label{eq:thermo_heat_responses}
\end{equation}
The derivatives are taken along $M(R)$ at fixed dimensional $\theta,
\eta$. Both quantities reduce to $-2\pi R^2$ in the Schwarzschild limit.
For the unexpanded pure KS geometry,
\begin{equation}
 C_A=C_M=-2\pi(R^2-\eta^2)=-8\pi M^2.
 \label{eq:thermo_heat_KS}
\end{equation}

\begin{figure*}[t]
 \centering
 \includegraphics[width=0.95\textwidth]{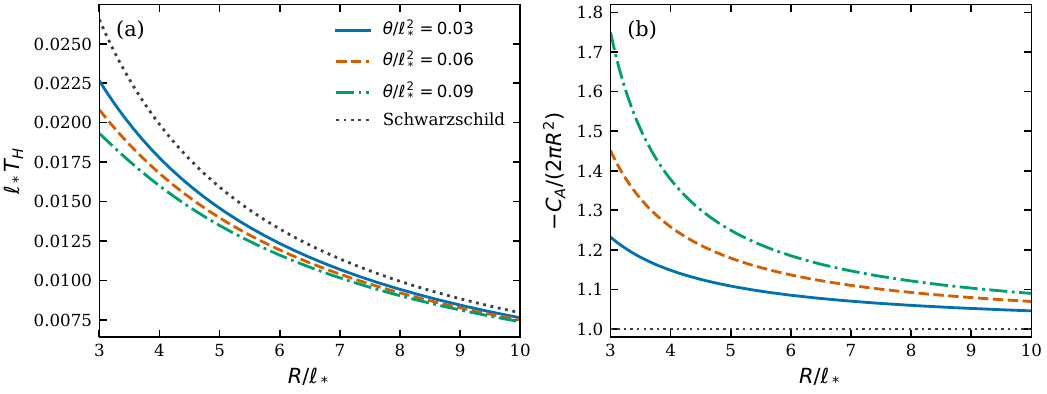}
 \caption{\rev{Thermodynamic quantities computed from the full exterior.
 A fixed reference length $\ell_*$ is used to compare horizon families
 with $\theta/\ell_*^2=0.03,0.06,0.09$ and
 $\eta/\ell_*=0.09$: (a) Hawking temperature and
 (b) the normalized entropy heat response. The dotted reference is
 Schwarzschild. Every displayed point belongs to an outer-horizon
 branch; the dimensional deformation parameters remain fixed along
 each curve as $M(R)$ varies.}}
 \label{fig:thermo_full}
\end{figure*}

Figure~\ref{fig:thermo_full} uses $3\leq R/\ell_*\leq10$.
The bound
$\mu(r)-r\mu'(r)\geq1-8w/(\pi r)>0$ throughout the retained
exterior ensures that $M(r)$ increases for $r\geq R$ and that each
plotted root is the outer horizon. On these branches the temperature
decreases monotonically and $C_A<0$. Increasing the smearing scale
lowers $T_H$ at fixed $R$ and increases the magnitude of the normalized
heat response. Both quantities approach their Schwarzschild behavior
as the horizon radius grows relative to the deformation scales.
\end{revision}

\begin{revision}
\section{\rev{Thin-shell construction and matter content}}
\label{sec:structure}
We join the exterior geometry of Sec.~\ref{sec1} to a de Sitter
interior across a timelike spherical shell. The two metrics are
\begin{equation}
 ds_\pm^2=-F_\pm(r_\pm)dt_\pm^2
 +\frac{dr_\pm^2}{F_\pm(r_\pm)}+r_\pm^2d\Omega_\pm^2,
 \label{eq:shell_metrics}
\end{equation}
where $F_+=F$ is the unexpanded exterior lapse and
\begin{equation}
 F_-(r)=1-H^2r^2,\qquad H^2=\frac{\Lambda_{\rm dS}}3>0,
 \qquad \alpha=H^{-1}.
 \label{eq:ds_interior}
\end{equation}
Thus $\alpha^2=3/\Lambda_{\rm dS}$. The interior vacuum has
$\rho_-=3H^2/(8\pi)$ and $p_-=-\rho_-$. We use the
Einstein--Israel junction conditions with the effective exterior source
specified by the metric~\cite{Israel,Visser:2003ge,Lobo:2015lbc}.

\subsection{\rev{Dynamical junction conditions}}

Let $r_\pm=a(\tau)$ describe the shell, with proper time $\tau$ and
induced metric
\begin{equation}
 ds_\Sigma^2=-d\tau^2+a(\tau)^2d\Omega^2.
 \label{eq:shell_induced}
\end{equation}
Define $\beta_\pm=\sqrt{F_\pm(a)+\dot a^2}$ and orient the common
normal from the interior toward increasing areal radius. The shell
four-velocity and normal are
\begin{equation}
 U_\pm^\mu=\left(\frac{\beta_\pm}{F_\pm},\dot a,0,0\right),
 \qquad n_\pm^\mu=\left(\frac{\dot a}{F_\pm},\beta_\pm,0,0\right),
 \label{normal}
\end{equation}
which satisfy $U^\mu U_\mu=-1$, $n^\mu n_\mu=1$, and $U^\mu n_\mu=0$.
With a prime denoting an areal-radius derivative, the independent
extrinsic-curvature components are
\begin{equation}
 K^\vartheta{}_{\vartheta\,\pm}=\frac{\beta_\pm}{a},
 \qquad
 K^\tau{}_{\tau\,\pm}=\frac{\ddot a+F_\pm'(a)/2}{\beta_\pm}.
 \label{eq:shell_curvature}
\end{equation}
For the jump $\kappa^i{}_j=K^i{}_{j+}-K^i{}_{j-}$,
the surface stress tensor $S^i{}_j={\rm diag}(-\sigma,\mathcal P,\mathcal P)$
obeys
\begin{equation}
 S^i{}_j=-\frac1{8\pi}
 \left(\kappa^i{}_j-\delta^i{}_j\kappa^k{}_k\right).
 \label{eq:shell_israel}
\end{equation}
It follows that
\begin{equation}
 \sigma=\frac{\beta_--\beta_+}{4\pi a},
 \label{gen-surfenergy2}
\end{equation}
\begin{align}
 \mathcal P=\frac1{8\pi a}\bigg[
 &\frac{F_++\dot a^2+a\ddot a+aF_+'/2}{\beta_+}
 \nonumber\\[-2pt]
 &-\frac{F_-+\dot a^2+a\ddot a+aF_-'/2}{\beta_-}
 \bigg].
 \label{gen-surfpressure2}
\end{align}
The combination entering the strong energy condition is
\begin{equation}
 \sigma+2\mathcal P=\frac1{4\pi}
 \left[\frac{\ddot a+F_+'/2}{\beta_+}
 -\frac{\ddot a-H^2a}{\beta_-}\right].
 \label{s2P}
\end{equation}
Both bulk geometries satisfy $\rho+p_r=0$, so the normal energy flux
vanishes and the conservation identity reduces to
\begin{equation}
 \frac{d}{d\tau}(4\pi a^2\sigma)
 +\mathcal P\frac{d}{d\tau}(4\pi a^2)=0.
 \label{eq:shell_conservation}
\end{equation}
This identifies the shell rest mass as $m_s=4\pi a^2\sigma$ and supplies
the matter conservation equation for a chosen shell equation of state.

\subsection{\rev{Static stresses and admissible radii}}

For an equilibrium radius $a_0$, we set $\dot a=\ddot a=0$ and require
\begin{equation}
 F_+(r)>0\quad(r\geq a_0),\qquad
 0<a_0<H^{-1},\qquad a_0>\eta.
 \label{eq:shell_domain}
\end{equation}
Whenever the exterior extension has horizons, the first condition places
the shell outside its largest root. The static density and pressure are
\begin{equation}
 \sigma_0=\frac{\sqrt{1-H^2a_0^2}-\sqrt{F_+(a_0)}}{4\pi a_0},
 \label{gen-surfenergy2a}
\end{equation}
\begin{align}
 \mathcal P_0=\frac1{8\pi a_0}\bigg[
 &\frac{F_+(a_0)+a_0F_+'(a_0)/2}{\sqrt{F_+(a_0)}}
 \nonumber\\[-2pt]
 &-\frac{1-2H^2a_0^2}{\sqrt{1-H^2a_0^2}}
 \bigg].
 \label{gen-surfpressure2a}
\end{align}
In particular, positivity of the density requires
\begin{equation}
 H^2\leq\frac{1-F_+(a_0)}{a_0^2}.
 \label{eq:shell_density_bound}
\end{equation}
The exterior lapse and its radial derivative enter independently in the
pressure, making the surface stresses sensitive to both the potential
and its gradient near the shell.

For fixed ADM mass, we use the dimensionless variables
\begin{align}
 x_0&=\frac{a_0}{M},&\lambda&=H^2M^2,&\hat\eta&=\frac{\eta}{M},
 \nonumber\\
 \Theta&=\frac{8\sqrt\theta}{\sqrt\pi M},&
 \widehat\sigma&=M\sigma,&\widehat{\mathcal P}&=M\mathcal P.
 \label{eq:shell_dimensionless}
\end{align}
The mass $M$ is the fixed normalization of the stresses, while
$\lambda$ fixes the interior curvature independently of $x_0$.
Writing $\mathsf F(x)=F_+(Mx)$ and a subscript $x$ for differentiation,
Eqs.~\eqref{gen-surfenergy2a} and~\eqref{gen-surfpressure2a} become
\begin{equation}
 \widehat\sigma_0=
 \frac{\sqrt{1-\lambda x_0^2}-\sqrt{\mathsf F(x_0)}}{4\pi x_0},
 \label{gen-surfenergyadm}
\end{equation}
\begin{align}
 \widehat{\mathcal P}_0=\frac1{8\pi x_0}\bigg[
 &\frac{\mathsf F(x_0)+x_0\mathsf F_x(x_0)/2}{\sqrt{\mathsf F(x_0)}}
 \nonumber\\[-2pt]
 &-\frac{1-2\lambda x_0^2}{\sqrt{1-\lambda x_0^2}}\bigg].
 \label{gen-surfpressureadm}
\end{align}

\begin{figure*}[t]
 \centering
 \includegraphics[width=0.98\textwidth]{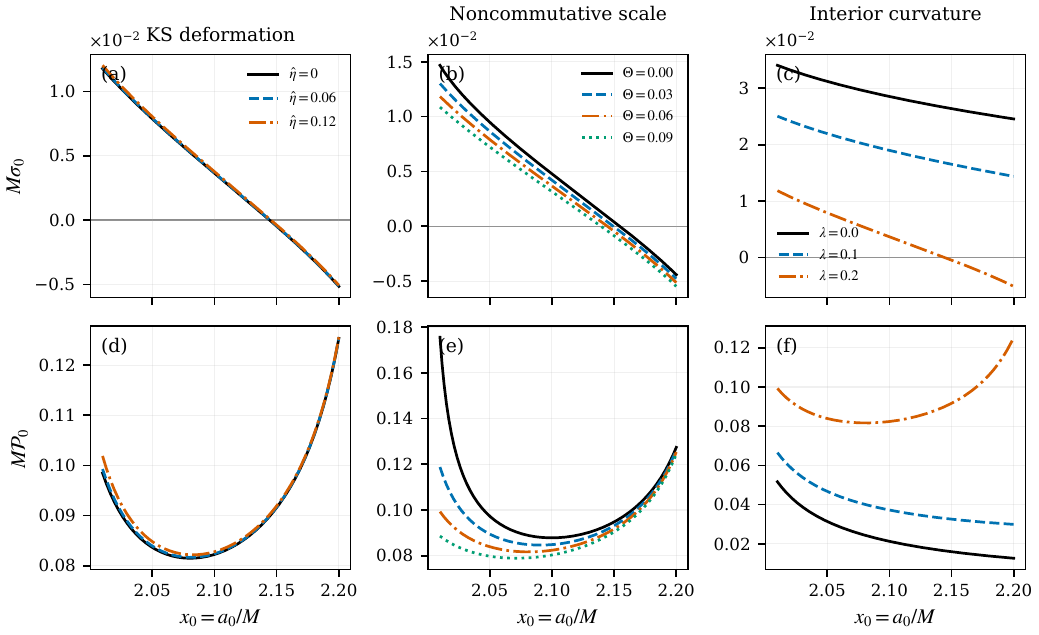}
 \caption{\rev{Static surface density (top row) and pressure
 (bottom row) calculated with the unexpanded exterior.
 The left column varies $\hat\eta=0,0.06,0.12$ at fixed
 $\Theta=0.06$, $\lambda=0.2$; the middle column varies
 $\Theta=0,0.03,0.06,0.09$ at fixed $\hat\eta=0.06$, $\lambda=0.2$;
 the right column varies $\lambda=0,0.1,0.2$ at fixed
 $\Theta=\hat\eta=0.06$. Each radius scan holds the two exterior
 parameters and the physical interior curvature fixed. The common
 interval $2.01\leq x_0\leq2.20$ lies outside every exterior outer
 horizon and inside each nonzero-curvature interior horizon.}}
 \label{fig:shell_parameters}
\end{figure*}

Figure~\ref{fig:shell_parameters} separates the three parameter
dependences. Increasing the KS scale lowers the exterior lapse and raises
the surface density. Increasing the noncommutative scale raises the
exterior lapse and lowers the density. Larger interior curvature reduces
$F_-(a_0)$, lowering $\sigma_0$ and increasing $\mathcal P_0$.
The inverse-square-root factors in Eq.~\eqref{gen-surfpressure2a}
produce rapid pressure growth near the simple exterior horizon and
the de Sitter horizon.

\subsection{\rev{Energy conditions of the shell}}

We evaluate the four-dimensional distributional shell stress
$T_{\hat\mu\hat\nu}^{\Sigma}={\rm diag}(\sigma,0,\mathcal P,\mathcal P)
\delta(\ell)$, where $\ell$ is proper distance normal to the shell.
The corresponding inequalities are
\begin{align}
 {\rm NEC/WEC}:\quad &\sigma\geq0,\quad\sigma+\mathcal P\geq0,
 \nonumber\\
 {\rm SEC}:\quad &\sigma\geq0,\quad\sigma+\mathcal P\geq0,
 \quad\sigma+2\mathcal P\geq0,
 \nonumber\\
 {\rm DEC}:\quad &\sigma\geq|\mathcal P|.
 \label{eq:shell_energy_conditions}
\end{align}
For null vectors tangent to the shell, the NEC reduces to
$\sigma+\mathcal P\geq0$; allowing a normal component adds
$\sigma\geq0$. These inequalities test the shell contribution separately
from the smooth interior and exterior stresses~\cite{Horvat:2007qa}.

\begin{figure*}[t]
 \centering
 \includegraphics[width=0.96\textwidth]{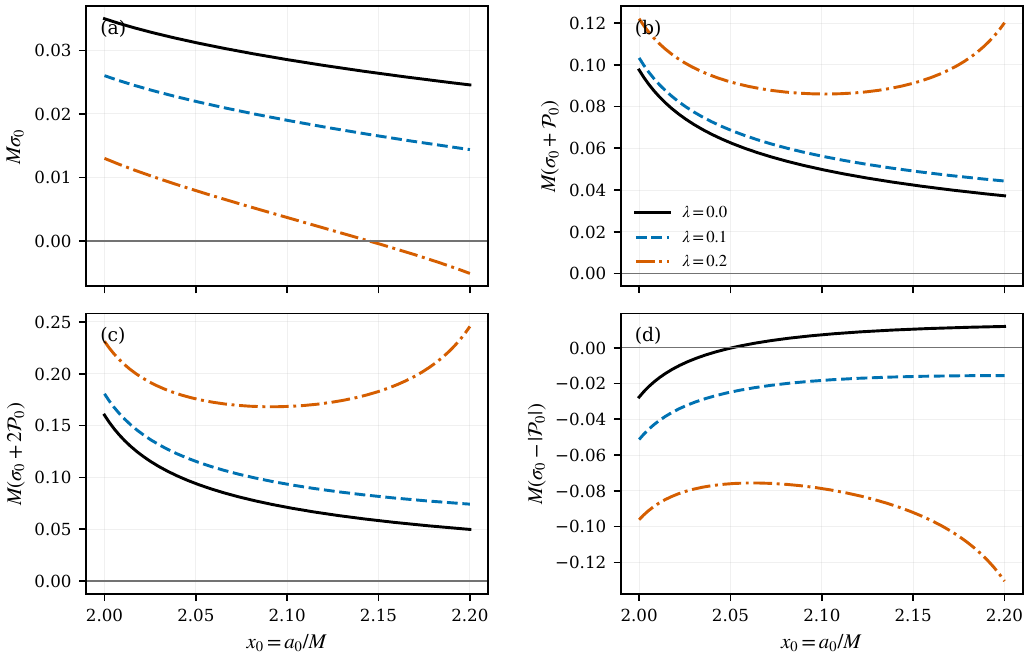}
 \caption{\rev{Energy-condition combinations for
 $\Theta=\hat\eta=0.06$ and fixed $\lambda=0,0.1,0.2$.
 The four panels display $M\sigma_0$, $M(\sigma_0+\mathcal P_0)$,
 $M(\sigma_0+2\mathcal P_0)$, and $M(\sigma_0-|\mathcal P_0|)$.
 The zero line makes the weak and dominant-condition boundaries explicit.
 All curves use the same unexpanded exterior and the mass normalization
 in Eq.~\eqref{eq:shell_dimensionless}.}}
 \label{fig:shell_energy_conditions}
\end{figure*}

The optical reference family $x_0=2$, $\Theta=0.06$, $\lambda=0.2$
has positive density and pressure for
$\hat\eta=0.03,0.06,0.09,0.12$. Across these four cases,
$M\sigma_0$ increases from $0.0129395$ to $0.0132232$, while
$M\mathcal P_0$ increases from $0.108220$ to $0.113281$.
The NEC, WEC, and SEC are satisfied at the shell, whereas the pressure
exceeds the density and violates the DEC.

The radius dependence is shown in Fig.~\ref{fig:shell_energy_conditions}.
For $\Theta=\hat\eta=0.06$ and $\lambda=0.2$, the density crosses zero
at $x_0=2.14464689$, setting the upper WEC boundary within the plotted
interval. The combinations $\sigma_0+\mathcal P_0$ and
$\sigma_0+2\mathcal P_0$ remain positive there.
For the flat-core limit $\lambda=0$, the DEC is recovered at
$x_0=2.05085820$ and remains satisfied over the remainder of the
displayed interval. The increase in interior curvature raises the
pressure relative to the density, tightening the allowed shell-matter
range. Radial stability can be studied by supplementing
Eq.~\eqref{eq:shell_conservation} with a response law
$d\mathcal P/d\sigma$ and perturbing the equilibrium radius.
\end{revision}

\section{\rev{Null geodesics and optical signatures}}
\label{sec:null_geodesics}
\begin{revision}
We consider a static shell with unit transmission, illuminated by a uniform
source plane at asymptotic infinity behind the object. A static observer
is placed at $D_o=2000M$. This extends the backlighting construction of
Refs.~\cite{Sakai:2014pga,Kubo:2016ada} to the full NC--KS exterior and
keeps the observer and source endpoints explicit. The primary numerical
family has its shell inside the outer photon orbit; a separate radius
scan follows the disappearance of the optical rings as the shell moves
beyond that orbit. In every case the retained exterior is static. We use the
unexpanded exterior $F_+(r)=\sqrt{1-\eta^2/r^2}-2\mathcal M_\theta(r)/r$
for the numerical geodesics, critical quantities, and images throughout this
section. The de Sitter interior has $F_-(r)=1-H^2r^2$.

\subsection{\rev{Geodesics and transparent-shell matching}}

In each static chart, spherical symmetry permits equatorial motion. With
$v$ an affine parameter, the Killing constants are
\begin{equation}
 E_\pm=F_\pm\frac{dt_\pm}{dv},\qquad
 L_\pm=r^2\frac{d\phi}{dv}.
 \label{eq:EL_definitions}
\end{equation}
The null constraint gives
\begin{equation}
 \begin{aligned}
 \left(\frac{dr}{dv}\right)^2&=E_\pm^2-\frac{F_\pm(r)L_\pm^2}{r^2},\\
 \left(\frac{dr}{d\phi}\right)^2
 &=r^4\frac{E_\pm^2}{L_\pm^2}-r^2F_\pm(r).
 \end{aligned}
 \label{eq:null_radial_equation}
\end{equation}
Continuity of the photon frequency and tangential momentum in the shell's
orthonormal frame fixes
\begin{equation}
 L_+=L_-=L,\qquad
 \frac{E_+}{\sqrt{F_+(a_0)}}=
 \frac{E_-}{\sqrt{F_-(a_0)}}.
 \label{eq:matching_EL}
\end{equation}
Define the redshift factor and exterior impact parameter by
\begin{equation}
 C=\frac{F_+(a_0)}{F_-(a_0)},\qquad
 b=\frac{L}{E_+}.
 \label{eq:C_matching}
\end{equation}
Then $E_-=E_+/\sqrt C$ and the interior impact parameter is
$b_-=\sqrt C\,b$. With the same ingoing or outgoing orientation, the
radial momentum satisfies
\begin{equation}
 \frac{(dr/dv)_+}{\sqrt{F_+(a_0)}}=
 \frac{(dr/dv)_-}{\sqrt{F_-(a_0)}}.
 \label{eq:matching_kr}
\end{equation}
The matched quantities describe transmission through the shell in a
single local inertial frame.

The de Sitter orbit integrates to
\begin{equation}
 r=r_m\sec(\phi-\phi_m),\qquad
 r_m=\left(\frac{1}{Cb^2}+H^2\right)^{-1/2}.
 \label{eq:interior_solution}
\end{equation}
For rays that enter the shell, the two crossings have angular separation
$2\arccos(r_m/a_0)$. The radial ray is obtained continuously as $b\to0$.
At infinity, the exterior trajectory approaches the straight-line form
$r=b\sec(\phi-\phi_c)$.

A common time coordinate, $t=t_+$ and $t_-=\sqrt C\,t$, expresses the
matched geometry as
\begin{equation}
 ds^2=-A(r)dt^2+B(r)dr^2+r^2d\Omega^2,
\end{equation}
with
\begin{equation}
 (A,B)=
 \begin{cases}
 (CF_-,F_-^{-1}),&r<a_0,\\
 (F_+,F_+^{-1}),&r\ge a_0.
 \end{cases}
 \label{eq:global_optical_metric}
\end{equation}
The global energy is $E=E_+$ and the radial constraint becomes
\begin{equation}
 AB\left(\frac{dr}{dv}\right)^2+L^2U(r)=E^2,
 \qquad U(r)=\frac{A(r)}{r^2}.
 \label{eq:continuous_potential}
\end{equation}
Thus $U(a_0^-)=U(a_0^+)=F_+(a_0)/a_0^2$. Its exterior local maximum
controls the critical orbit; its central behavior is $U\sim C/r^2$.
The interior derivative $U_-'=-2C/r^3$ is negative. For
$a_0<r_{\rm ph}$, the exterior derivative at the shell is positive,
so the matched potential has a cusp minimum followed by an exterior
maximum, as in Ref.~\cite{Sakai:2014pga}. Moving the shell beyond
$r_{\rm ph}$ removes that exterior maximum from the physical spacetime.

\begin{table}[tbp]
\revisioncolor
\centering
\caption{\rev{Critical optical quantities from the unexpanded exterior for
$\Theta=0.06$, $x_0=2$, and $\lambda=0.20$.}}
\label{tab:critical_optics}
\begin{ruledtabular}
\begin{tabular}{cccc}
$\widehat\eta$ & $C$ & $r_{\rm ph}/M$ & $b_c/M$\\
0.03 & 0.074431 & 2.959763 & 5.143971\\
0.06 & 0.072743 & 2.960688 & 5.145168\\
0.09 & 0.069928 & 2.962228 & 5.147162\\
0.12 & 0.065985 & 2.964383 & 5.149951\\
\end{tabular}
\end{ruledtabular}
\end{table}

\begin{figure*}[tp]
\centering
\includegraphics[width=0.98\textwidth]{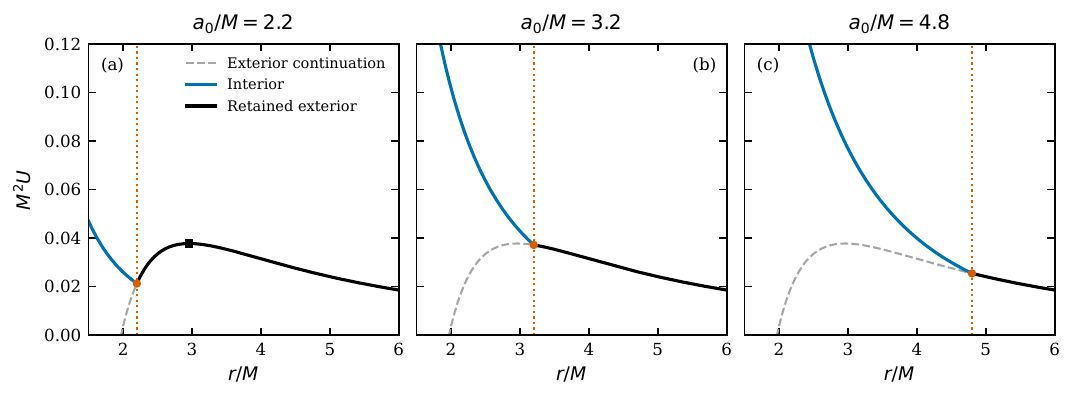}
\caption{\rev{Matched optical potential for $\Theta=\hat\eta=0.06$ and
$\lambda=0.01$, with increasing shell radius. Blue and black curves
show the interior and retained exterior; the gray dashed curve shows
the continuation of the exterior formula below the shell. Orange
markers and vertical lines identify the shell. The black square in
panel (a) marks the exterior maximum. Panels (b) and (c) place the
shell beyond that maximum.}}
\label{fig:shadow_potential}
\end{figure*}

\subsection{\rev{Critical orbit and numerical trajectories}}

In the exterior, $u=1/r$ obeys
\begin{align}
 \left(\frac{du}{d\phi}\right)^2&=b^{-2}-u^2F_+(1/u),\label{eq:u_general}\\
 \frac{d^2u}{d\phi^2}&=-uF_+(1/u)+\frac12F_+'(1/u).
 \label{eq:u_exact}
\end{align}
Here $F_+'$ denotes differentiation with respect to the areal radius.
The leading inverse-radius metric reduces the second equation to
$u''=-u+3Mu^2-2qu^3$, where
$q=8M\sqrt{\theta/\pi}-\eta^2/2$.
For the unexpanded exterior we solve
\begin{equation}
 \begin{gathered}
 r_cF_+'(r_{\rm ph})-2F_+(r_{\rm ph})=0,\\
 U''(r_{\rm ph})<0,\qquad b_c=\frac{r_{\rm ph}}{\sqrt{F_+(r_{\rm ph})}}.
 \end{gathered}
 \label{eq:bc_exact}
\end{equation}
The analytic leading-order comparison is
\begin{equation}
 r_{\rm ph}^{(2)}=\frac{3M+\sqrt{9M^2-8q}}2.
 \label{eq:rph_approx}
\end{equation}
The entry threshold depends on whether the photon barrier is retained:
\begin{equation}
 b_{\rm ent}=\begin{cases}
 b_c,&a_0\leq r_{\rm ph},\\
 b_\Sigma\equiv a_0/\sqrt{F_+(a_0)},&a_0>r_{\rm ph}.
 \end{cases}
 \label{eq:entry_threshold}
\end{equation}
Incoming rays with $b<b_{\rm ent}$ cross the shell. For
$b>b_{\rm ent}$, the outermost accessible solution of
\begin{equation}
 b^{-2}=F_+(r_t)/r_t^2
 \label{eq:turning_radius}
\end{equation}
sets the exterior turning point, with $r_t\geq a_0$.
For $a_0<r_{\rm ph}$, a ray with $b=b_c$ asymptotically approaches
$r_{\rm ph}$. For $a_0>r_{\rm ph}$, the threshold ray grazes the shell at
$r_t=a_0$.

For the numerical family we fix
\begin{equation}
 \begin{gathered}
 \Theta=0.06,\qquad x_0=2,\qquad\lambda=0.20,\\
 \widehat\eta=0.03,\ 0.06,\ 0.09,\ 0.12,
 \end{gathered}
 \label{eq:optical_parameters}
\end{equation}
using the mass-normalized variables defined above. In particular,
$F_-(a_0)=1-\lambda x_0^2=0.20$.
The shell lapse ranges from $0.0148861$ to $0.0131970$ and remains
positive. For this family, the bound
$\mu(x)-x\mu'(x)\ge1-\Theta/x>0$ for $x\ge x_0$
ensures that the exterior lapse increases throughout the retained region.
Table~\ref{tab:critical_optics} reports the critical quantities from the
full lapse. Increasing $\widehat\eta$ lowers the barrier and moves its
maximum and critical impact parameter outward.

\begin{figure*}[tp]
\centering
\includegraphics[width=\textwidth]{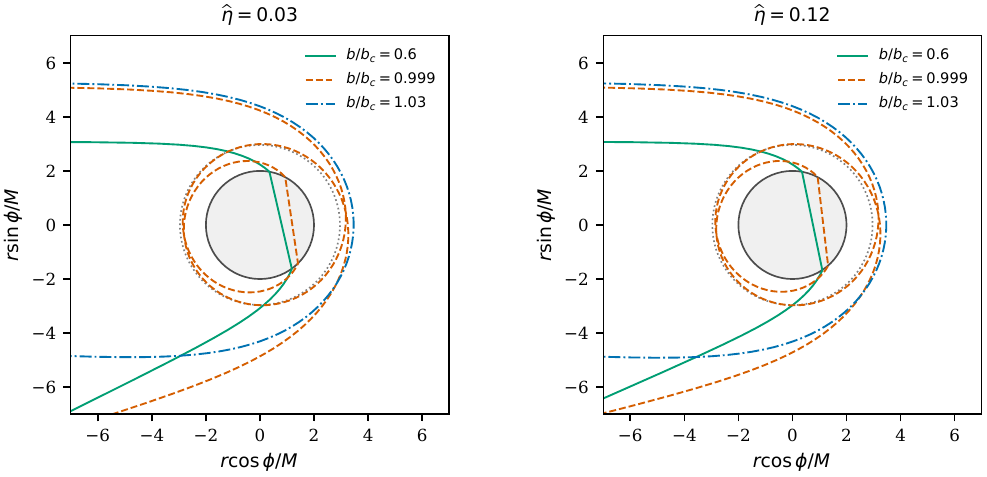}
\caption{\rev{Null trajectories in the areal-radius plane for
$\Theta=0.06$, $x_0=2$, and $\lambda=0.20$, with
$\widehat\eta=0.03$ (left) and $0.12$ (right).
These trajectories illustrate the asymptotic-observer limit.
The shaded disk ends at the transparent shell; the dotted circle marks
$r_{\rm ph}$. The rays have $b/b_c=0.6$ (solid), $0.999$ (dashed), and
$1.03$ (dash-dotted). Penetrating rays traverse the interior, while the
exterior-return ray turns outside the photon orbit.}}
\label{fig:optical_trajectories}
\end{figure*}

\subsection{\rev{Observer, deflection and source-plane visibility}}

At the static observer, let $\psi$ be the angle between a backward-traced
ray and the inward radial direction. Its relation to the conserved
impact parameter is
\begin{equation}
 \sin\psi=\frac{b\sqrt{F_+(D_o)}}{D_o},\qquad
 \rho_{\rm cam}=D_o\tan\psi.
 \label{eq:camera_angle}
\end{equation}
The plots use the calibrated impact coordinates $(b_y,b_z)$, with
$b^2=b_y^2+b_z^2$. Equation~\eqref{eq:camera_angle} converts their
radial scale to a physical angular image for the specified observer.

Define the exterior angular integral
\begin{equation}
 I(r_1,r_2;b)=\int_{r_1}^{r_2}
       \frac{b\,dr}{r^2\sqrt{1-b^2F_+(r)/r^2}}.
 \label{eq:angular_integral}
\end{equation}
For a ray traced backward from the observer through the shell to the
asymptotic source side, the deflection relative to the opposite radial
direction is
\begin{equation}
 \begin{aligned}
 \Phi_<(b)={}&I(a_0,D_o;b)+I(a_0,\infty;b)\\
 &+2\arccos(r_m/a_0)-\pi,
 \qquad b<b_{\rm ent}.
 \end{aligned}
 \label{eq:Phi_crossing}
\end{equation}
For an exterior-return ray,
\begin{equation}
 \Phi_>(b)=I(r_t,D_o;b)+I(r_t,\infty;b)-\pi,
 \label{eq:Phi_turning}
\end{equation}
with $b>b_{\rm ent}$. The radial limit is $\Phi_<(0)=0$.
Relative to the phase $\Phi_\infty$ with both endpoints at infinity,
the finite observer subtracts the smooth tail
\begin{align}
 J_D(b)&=I(D,\infty;b)
       =\frac bD+O\!\left(\frac{b^3}{D^3}\right),\nonumber\\
 \Phi(b)&=\Phi_\infty(b)-J_{D_o}(b).
 \label{eq:finite_observer_tail}
\end{align}
Cutting both endpoints off at the same radius $D_o$ instead gives
$\Phi_{D_o,D_o}=\Phi_\infty-2J_{D_o}$. We use this second convention
only for the numerical benchmark of Ref.~\cite{Sakai:2014pga}.

For the source plane behind the object, the asymptotic outgoing
direction of a backward ray has negative longitudinal component
when $\cos\Phi>0$. Its geometrical visibility is therefore
\begin{equation}
 \mathcal V_{\rm GS}(b)
 =\frac{1+\operatorname{sgn}[\cos\Phi(b)]}{2}.
 \label{eq:visibility_mask}
\end{equation}
The bright--dark boundaries solve
\begin{equation}
 \Phi(b)=\left(n-\frac12\right)\pi,\qquad n=1,2,\ldots.
 \label{eq:optical_boundaries}
\end{equation}
We label penetrating roots by $\zeta_n$ and exterior-return roots
by $\omega_n$. The central bright disk has $0\leq b<\zeta_1$,
followed by the dark annulus $\zeta_1<b<\zeta_2$. For
$a_0<r_{\rm ph}$, further alternating annuli accumulate on $b_c$.
These dark regions are defined by the prescribed rear illumination;
their null rays return toward the observer-side sky, as discussed
in Ref.~\cite{Sakai:2014pga}.

For comparison, the associated horizon-bearing exterior has the
same $\Phi_>(b)$ but captures rays below $b_c$. Its backlit image is
\begin{equation}
 \mathcal V_{\rm BH}(b)=
 \begin{cases}
 0,&b<b_c,\\
 [1+\operatorname{sgn}(\cos\Phi_>(b))]/2,&b>b_c.
 \end{cases}
 \label{eq:BH_visibility}
\end{equation}
For $a_0<r_{\rm ph}$, the gravastar and black-hole maps coincide
outside the critical circle under the same observer and illumination
conditions. The
penetrating trajectories generate the additional central disk and
inner annuli in the gravastar image.

\subsection{\rev{Schwarzschild benchmark and deformed shadow}}

The reference uses $r_g=2M$ and $h=r_gH$. Its $h=0.4$ stiff-shell
equilibrium, $\mathcal P_0=\sigma_0$, gives
$a_0/r_g=1.3027244501$ and $C=0.3189965465$, in agreement with the
rounded configuration used in Ref.~\cite{Sakai:2014pga}.
With $\Theta=\eta=0$, the photon orbit and critical scale are
$r_{\rm ph}/r_g=3/2$ and $b_c/r_g=3\sqrt3/2$.
Table~\ref{tab:shadow_benchmark} gives the boundary values obtained
directly from the geodesic equations for the three endpoint
conventions. The first column reproduces the finite-radius
construction underlying the reference deflection plot. The other
columns quantify the effect of the observer and source limits.

\begin{table}[tbp]
\revisioncolor
\centering
\caption{\rev{Schwarzschild--de Sitter benchmark at $h=0.4$ and
$a_0/r_g=1.3027244501$. Finite endpoints are at $D=1000r_g$;
all boundary lengths are divided by $r_g$.}}
\label{tab:shadow_benchmark}
\begin{ruledtabular}
\begin{tabular}{lccc}
 & $(D,D)$ & $(D,\infty)$ & $(\infty,\infty)$\\
$\zeta_1/r_g$ & 1.73144784 & 1.73036009 & 1.72927227\\
$\zeta_2/r_g$ & 2.48851742 & 2.48835657 & 2.48819549\\
$\omega_1/r_g$ & 3.07991760 & 3.08184462 & 3.08378338\\
$\omega_2/r_g$ & 2.61388524 & 2.61392722 & 2.61396930\\
\end{tabular}
\end{ruledtabular}
\end{table}

\begin{figure*}[tp]
\centering
\includegraphics[width=\textwidth]{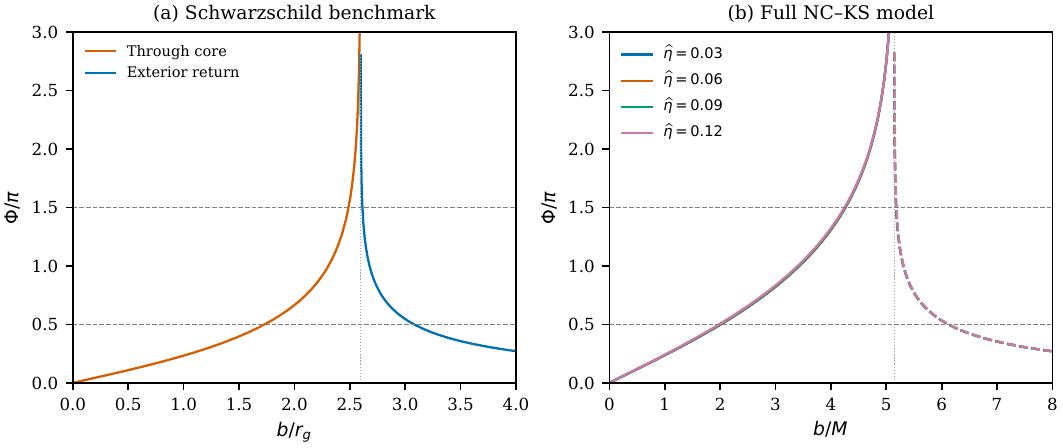}
\caption{\rev{Deflection and visibility-boundary construction.
Left: the Schwarzschild--de Sitter reference configuration with
$h=0.4$, $a_0/r_g=1.3027244501$ and both endpoints at $1000r_g$.
Right: the full NC--KS geometries of Eq.~\eqref{eq:optical_parameters}
with an observer at $2000M$ and an asymptotic source plane.
Penetrating and exterior-return branches approach the critical
impact parameter from opposite sides; in the right panel they are
solid and dashed, respectively. Horizontal lines identify
the first two phase thresholds for the visibility boundaries.}}
\label{fig:shadow_deflection}
\end{figure*}

For the NC--KS family, Table~\ref{tab:potential_values} gives the
finite-observer boundaries and the exterior potential maximum.
Increasing $\hat\eta$ moves the first three penetrating boundaries
inward and the exterior-return boundaries outward.
Across the tabulated family, the central bright radius decreases
by $1.66\%$, while $b_c$ increases by $0.116\%$.
For $\hat\eta=0.06$, placing the observer at $2000M$ instead of
infinity changes $\zeta_1/M$ from $2.00744857$ to $2.00856902$
and $\omega_1/M$ from $6.12377444$ to $6.11991137$.
The finite-distance correction is therefore resolved separately
from the deformation dependence.

\begin{table*}[t]
\revisioncolor
\centering
\caption{\rev{Full-metric potential maximum and visibility boundaries
for $\Theta=0.06$, $a_0/M=2$, $\lambda=0.20$, $D_o/M=2000$,
and the asymptotic source plane. Boundary lengths are in units of $M$.}}
\label{tab:potential_values}
\begin{ruledtabular}
\begin{tabular}{ccccccc}
$\widehat\eta$ & $M^2U_{\max}$ & $\zeta_1/M$ & $\zeta_2/M$
 & $\zeta_3/M$ & $\omega_2/M$ & $\omega_1/M$\\
0.03 & 0.03779227 & 2.015127 & 4.264485 & 4.936328 & 5.176526 & 6.118888\\
0.06 & 0.03777469 & 2.008569 & 4.261059 & 4.936267 & 5.177702 & 6.119911\\
0.09 & 0.03774542 & 1.997507 & 4.255156 & 4.936103 & 5.179663 & 6.121617\\
0.12 & 0.03770455 & 1.981730 & 4.246464 & 4.935733 & 5.182405 & 6.124002\\
\end{tabular}
\end{ruledtabular}
\end{table*}

\begin{figure*}[tp]
\centering
\includegraphics[width=0.92\textwidth]{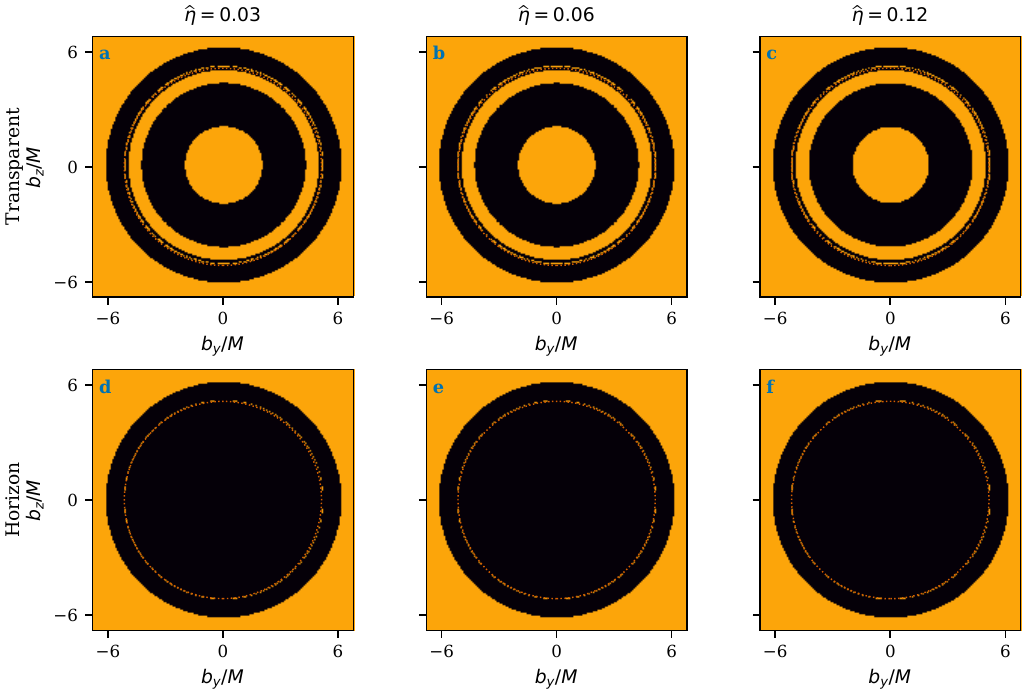}
\caption{\rev{Backlit visibility maps of the transparent gravastar
(top) and its associated black-hole exterior (bottom).
Columns use $\hat\eta=0.03,0.06,0.12$, with the remaining parameters
as in Table~\ref{tab:potential_values}. Amber and black denote
visible and dark regions of the same uniform asymptotic source plane.
The dotted critical circle marks $b_c$; the axes use calibrated impact
coordinates. Each map samples $[-6.8M,6.8M]^2$ on an
$800\times800$ grid. Eight penetrating and four exterior-return
boundaries resolve the annuli down to an accumulation interval
smaller than $9.24\times10^{-5}M$, compared with the $0.017M$
pixel width.}}
\label{fig:optical_visibility}
\end{figure*}

We evaluate the angular integrals in $u=1/r$, using
$u=u_t(1-t^2)$ to regularize an exterior turning point.
Six independent 50-digit quadratures, including the Schwarzschild
benchmark and NC--KS roots, give phase residuals below
$3.2\times10^{-12}$ radians. Tightening the numerical tolerances
by a factor of ten changes the tested impact parameters by less
than $4.6\times10^{-14}M$.

\subsection{\rev{Ring mergers as the shell radius increases}}
\label{sec:radius_mergers}

To follow the shell across the photon orbit while keeping its
interior curvature fixed, we consider a second family,
\begin{equation}
 \Theta=\hat\eta=0.06,\qquad\lambda=0.01,
 \qquad2.01\leq a_0/M\leq5.30.
 \label{eq:radius_scan_family}
\end{equation}
The entire scan lies outside the exterior horizon and below the
de Sitter horizon $H^{-1}=10M$. The surface density is positive
throughout. The reference's $\mathcal P=\sigma$ condition fixes
its benchmark shell; the present scan varies the equilibrium
stresses according to Eqs.~\eqref{gen-surfenergy2a} and
\eqref{gen-surfpressure2a} at fixed $H^2$.

For $a_0>r_{\rm ph}$, the penetrating phase is finite. Numerically,
it increases toward its grazing maximum in this family. Since
$r_m(b_\Sigma)=a_0$, its interior angular contribution vanishes
there, and
\begin{equation}
 \Phi_{\max}(a_0)=I(a_0,D_o;b_\Sigma)
                   +I(a_0,\infty;b_\Sigma)-\pi.
 \label{eq:grazing_phase}
\end{equation}
A penetrating boundary and its exterior partner merge when the
grazing phase reaches the corresponding threshold. Equivalently,
\begin{equation}
 a_{0,n}=r_t(\omega_n),\qquad
 b_\Sigma(a_{0,n})=\zeta_n=\omega_n.
 \label{eq:merger_condition}
\end{equation}
This relation generalizes the geometric construction in
Ref.~\cite{Sakai:2014pga} to the deformed exterior. It also explains
why the exterior boundary and its merger radius depend on the
exterior metric and endpoint prescription, while the approach of
the penetrating boundary depends on the interior.

\begin{figure*}[tp]
\centering
\includegraphics[width=\textwidth]{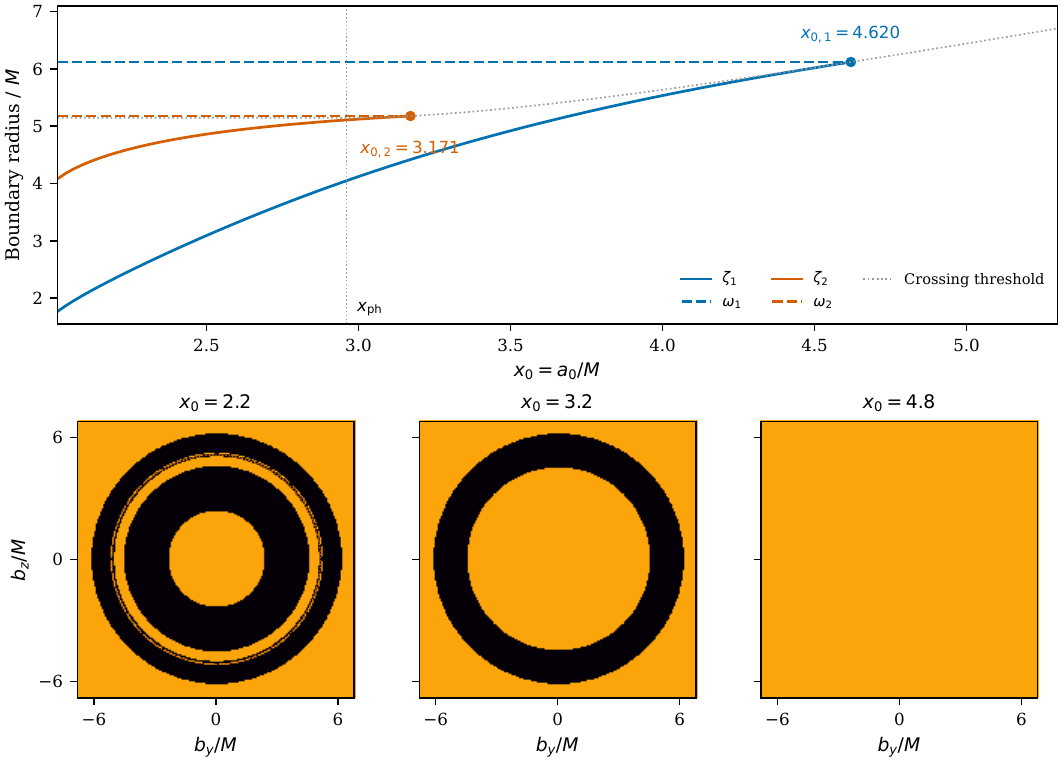}
\caption{\rev{Evolution of the backlit rings with shell radius for
Eq.~\eqref{eq:radius_scan_family}, $D_o/M=2000$, and the asymptotic
source plane. Amber marks visible source-plane pixels and black
marks dark pixels. The boundary curves end where the penetrating and
exterior-return roots merge. The vertical line marks
$x_{\rm ph}=r_{\rm ph}/M$. The lower maps show the same family
at $a_0/M=2.2,3.2,4.8$. At $3.2M$ one dark annulus remains;
at $4.8M$ the source plane is visible throughout the displayed image.}}
\label{fig:shadow_radius_scan}
\end{figure*}

The exterior photon orbit is at $r_{\rm ph}/M=2.96068756$.
The second boundary pair merges at $a_{0,2}/M=3.17092181$,
where $\zeta_2/M=\omega_2/M=5.17770248$.
The first pair merges at $a_{0,1}/M=4.61956378$,
where $\zeta_1/M=\omega_1/M=6.11991137$.
Between these radii, a single dark annulus lies between
$\zeta_1$ and $\omega_1$.
At $a_0/M=4.8$, the maximum phase is $1.46082635<\pi/2$,
giving the uniformly visible map in Fig.~\ref{fig:shadow_radius_scan}.
Thus the dark annulus persists over a finite range of shell radii
after the exterior photon orbit has been excised.
\end{revision}

\section{\rev{Thin-disk transfer functions and intensity maps}}
\label{sec:disk_images}
\begin{revision}
The optical appearance also depends on the distribution of emitting
matter. We now place a geometrically and optically thin test layer in
the equatorial plane and view it from the symmetry axis at $D_o=2000M$.
The layer emits isotropically in the local static frame. We neglect
absorption and take unit transmission through the shell. This construction adapts the
transfer-function method of Refs.~\cite{GrallaHolzWald2019,PengGuoFengWormhole2021}
to the de Sitter interior and the full NC--KS exterior. We use the
same exterior emission profile for the gravastar and its associated
black-hole geometry.

\subsection{\rev{Disk crossings and redshift}}

Let $\chi$ denote the accumulated orbital angle of a backward-traced
ray, measured from the observer's polar axis. Equatorial crossings
occur at
\begin{equation}
 \chi_m=\left(m-\frac12\right)\pi,\qquad m=1,2,\ldots.
 \label{eq:disk_crossing_angles}
\end{equation}
The transfer function $\mathcal R_m(b)$ is the areal radius at the
$m$th crossing. On the incoming exterior leg,
$\chi(r)=I(r,D_o;b)$. For a penetrating ray, write
$\chi_\Sigma=I(a_0,D_o;b)$ and $\alpha=\arccos(r_m/a_0)$,
where $r_m$ is the interior turning radius in
Eq.~\eqref{eq:interior_solution}. The ray exits the core at
$\chi_\Sigma+2\alpha$, and its outgoing exterior angle is
\begin{equation}
 \chi_{\rm out}(r)=\chi_\Sigma+2\alpha+I(a_0,r;b).
 \label{eq:disk_outgoing_angle}
\end{equation}
An exterior-return ray instead has
$\chi_{\rm out}(r)=I(r_t,D_o;b)+I(r_t,r;b)$.
Solving these relations at the angles in
Eq.~\eqref{eq:disk_crossing_angles} gives all exterior transfer
branches. We assign emission to crossings within the disk's radial
support. A captured black-hole ray contributes at any such crossing
before reaching the horizon.

For a static emitter at radius $r$, the frequency ratio measured by
the static observer is
\begin{equation}
 g(r)=\frac{\nu_o}{\nu_e}
     =\sqrt{\frac{F_+(r)}{F_+(D_o)}}.
 \label{eq:disk_redshift}
\end{equation}
Conservation of $I_\nu/\nu^3$ along a ray gives a factor $g^3$ in
the specific intensity. Integrating over observed frequency supplies
one further factor of $g$. If $\mathcal J(r)$ is the bolometric
intensity added by one disk crossing in the static emitting frame,
the observed intensity is
\begin{equation}
 I_o(b)=\sum_{m\in\mathcal E(b)}
 \left[\frac{F_+(\mathcal R_m(b))}{F_+(D_o)}\right]^2
 \mathcal J(\mathcal R_m(b)),
 \label{eq:disk_intensity}
\end{equation}
where $\mathcal E(b)$ contains the exterior crossings. The core and
shell have zero emissivity. The matching factor $C$ affects the
crossing radii through the interior trajectory, while the endpoint
frequency shift follows Eq.~\eqref{eq:disk_redshift}.

For the shared exterior disk with inner edge above $a_0$, the two
geometries have identical incoming contributions. The gravastar adds
the outgoing contributions after transmission through the core.
Consequently, for $a_0<r_{\rm ph}$,
\begin{equation}
 \begin{aligned}
 I_o^{\rm GS}(b)&\geq I_o^{\rm BH}(b),&& b<b_c,\\
 I_o^{\rm GS}(b)&=I_o^{\rm BH}(b),&& b>b_c.
 \end{aligned}
 \label{eq:disk_comparison}
\end{equation}
These relations provide direct checks of the transfer calculation.
The radial ray has no exterior equatorial crossing, so $I_o(0)=0$
for both objects. Thus the same transparent core that produces a
bright center under rear-plane illumination can have a dark center
when illuminated by an exterior disk.

\subsection{\rev{Emission profiles and color images}}

We evaluate the images at $\Theta=\hat\eta=0.06$, $a_0/M=2$, and
$\lambda=0.20$. Two prescribed profiles isolate the effect of moving
the emitting inner edge inward. The first is concentrated near the
outer marginally stable circular orbit, whose radius solves
\begin{equation}
 r F_+F_+''+3F_+F_+'-2r(F_+')^2=0,
 \qquad r=r_{\rm ISCO}.
 \label{eq:disk_isco}
\end{equation}
The outer solution is selected on the timelike circular-orbit branch,
with $2F_+-rF_+'>0$. The second profile extends inward to the photon
orbit. For the chosen exterior, $r_{\rm ISCO}/M=5.91178880$ and
$r_{\rm ph}/M=2.96068756$. We choose
\begin{align}
 \frac{\mathcal J_{\rm I}(r)}{I_*}
 &=e^{-(r-r_{\rm ISCO})/M}\,\mathsf H(r-r_{\rm ISCO}),
 \label{eq:disk_emission_I}\\
 \frac{\mathcal J_{\rm II}(r)}{I_*}
 &=\frac{1+e^{-4}}{1+e^{2(r-r_{\rm ph}-2M)/M}}
       \,\mathsf H(r-r_{\rm ph}),
 \label{eq:disk_emission_II}
\end{align}
where $\mathsf H$ is the step function and $I_*$ sets the emitted
intensity scale. These explicit profiles follow the sharply truncated
and extended inner-emission patterns illustrated in
Refs.~\cite{GrallaHolzWald2019,PengGuoFengWormhole2021}.
Here $r_{\rm ISCO}$ specifies an emission edge; the prescribed
emitting frame remains static in both models.
Figure~\ref{fig:disk_transfer} displays the emission profiles and
the first four crossing functions. Figures~\ref{fig:disk_model_I}
and \ref{fig:disk_model_II} compare the corresponding intensity
profiles and images at the same observer distance and normalization.

\begin{figure*}[tp]
\centering
\includegraphics[width=\textwidth]{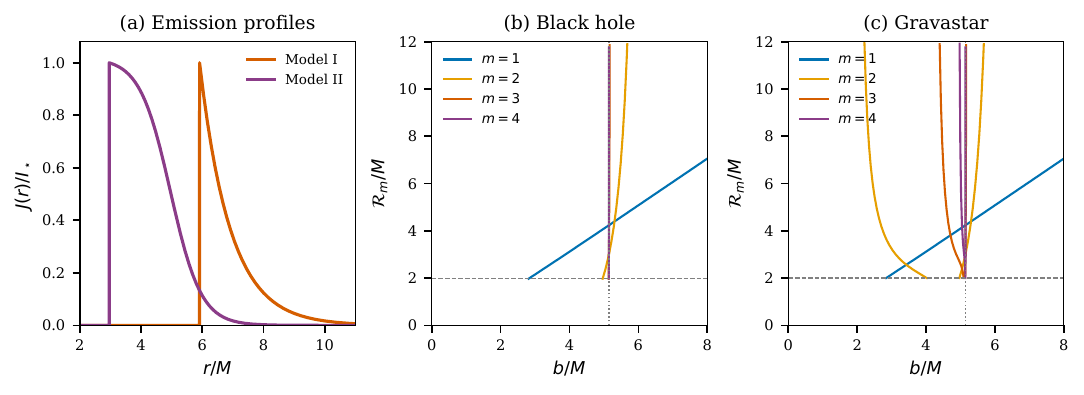}
\caption{\rev{Emission profiles and face-on disk transfer functions for
$\Theta=\hat\eta=0.06$, $a_0/M=2$, $\lambda=0.20$, and $D_o/M=2000$.
The transfer curves show the first four exterior equatorial crossings
of backward rays, retaining their chronological indices. The gravastar includes additional outgoing branches after core
transmission. The critical impact parameter separates transmitted
and exterior-return trajectories and is marked by the vertical dotted
line. Horizontal dashed lines mark the horizon in the black-hole
panel and the shell in the gravastar panel.}}
\label{fig:disk_transfer}
\end{figure*}

\begin{figure*}[tp]
\centering
\includegraphics[width=\textwidth]{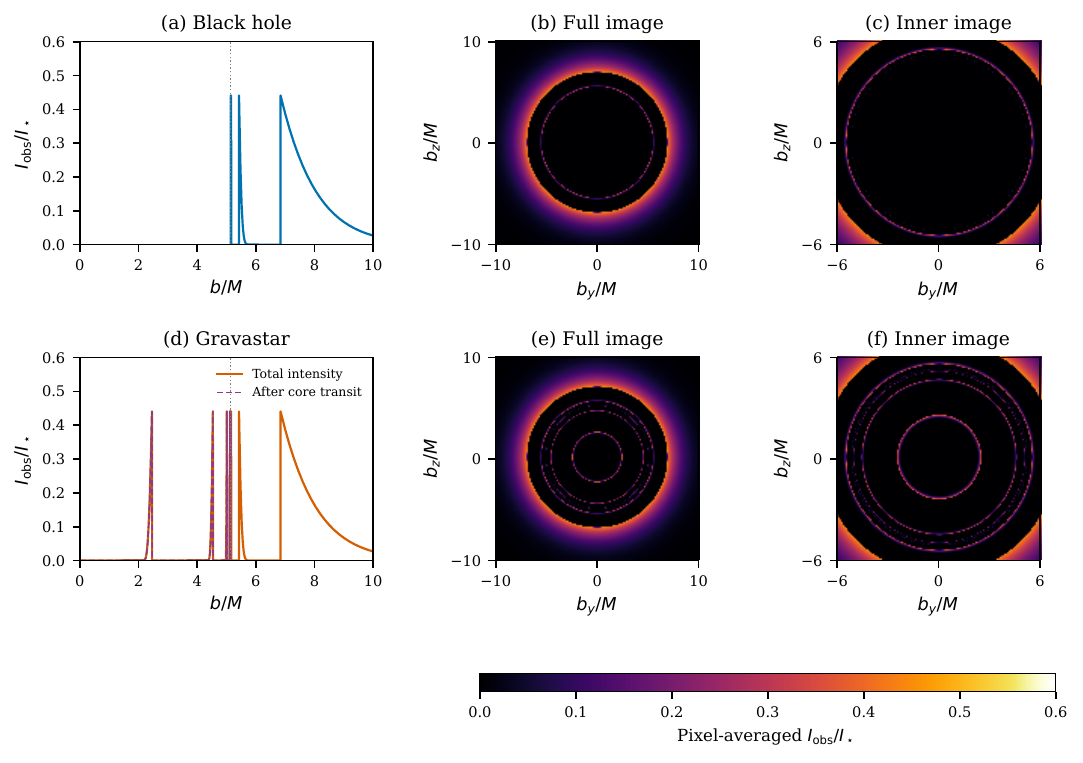}
\caption{\rev{Observed bolometric intensity for emission model I.
The black-hole and transparent-gravastar rows use the same exterior,
disk profile and observer. Columns show the radial intensity,
the full image, and the inner image. The dashed purple curve isolates
the light received after core transit. The maps use a common linear
scale $0\leq\overline I_o/I_*\leq0.6$, also used in
Fig.~\ref{fig:disk_model_II}; purple through orange, yellow and white
indicate increasing intensity. Each $800\times800$ map averages
the intensity over its pixels. The full and inner fields span
$[-10M,10M]^2$ and $[-6M,6M]^2$, respectively. Image coordinates are
the calibrated impact coordinates of Eq.~\eqref{eq:camera_angle}.}}
\label{fig:disk_model_I}
\end{figure*}

\begin{figure*}[tp]
\centering
\includegraphics[width=\textwidth]{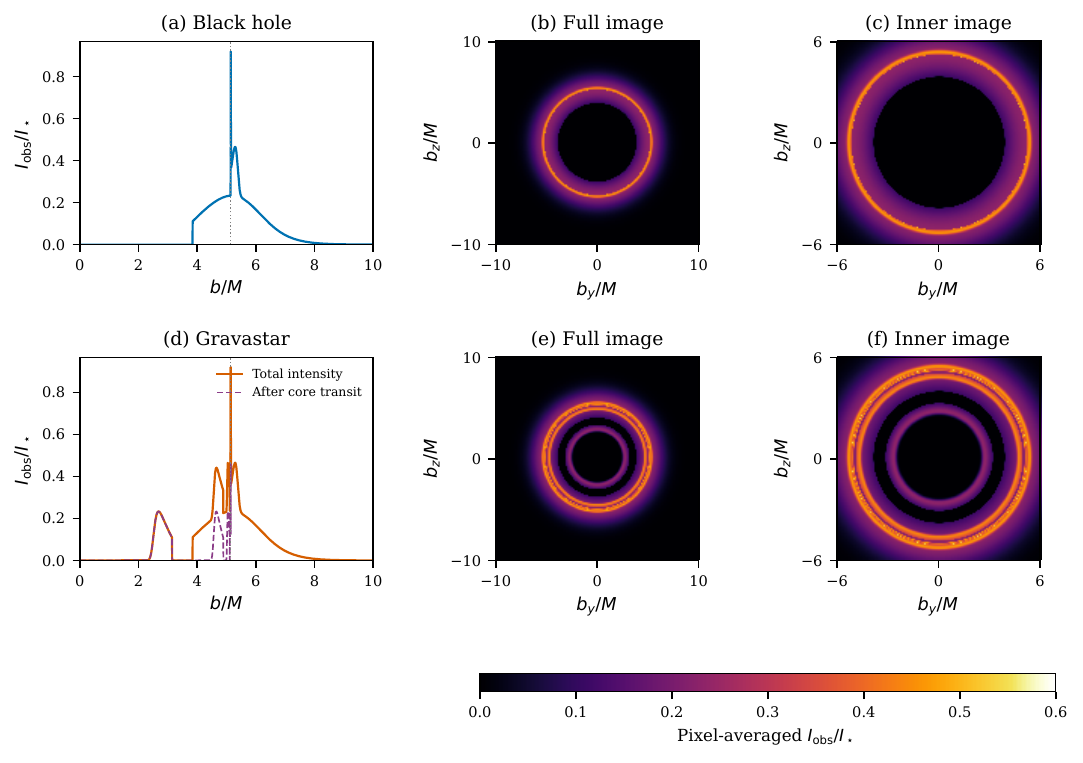}
\caption{\rev{The same comparison for emission model II, which extends
to $r_{\rm ph}$. Contributions from repeated disk crossings produce
additional inner structure in the transparent-gravastar image.
The color scale represents the same bolometric quantity and is shared
by both geometries and both emission models. The radial curves show
pointwise samples, while the image colors show pixel averages.}}
\label{fig:disk_model_II}
\end{figure*}

The color maps represent the intensity in Eq.~\eqref{eq:disk_intensity}.
The emission profile and redshift set each crossing's intensity;
the slope $d\mathcal R_m/db$ controls how its emitting radial range
maps into an apparent ring width.
The critical circle continues to be fixed by the exterior geometry.
For emission extending to $r_{\rm ph}$, successively higher crossing
orders accumulate near $b_c$. The ideal critical ray has an unbounded
number of crossings in this optically thin prescription; finite image
pixels integrate the resulting logarithmic intensity enhancement.
For a pixel $P$ of area $A_P$ in the calibrated impact plane we compute
\begin{equation}
 \overline I_P=\frac1{A_P}\int_P
 I_o\!\left(\sqrt{b_y^2+b_z^2}\right)\,db_y\,db_z.
 \label{eq:disk_pixel_average}
\end{equation}
The angular measure of a circular image annulus follows from
Eq.~\eqref{eq:camera_angle}:
\begin{equation}
 d\Omega_o=\frac{2\pi F_+(D_o)b\,db}
 {D_o^2\sqrt{1-F_+(D_o)b^2/D_o^2}}.
 \label{eq:disk_solid_angle}
\end{equation}
We integrate $I_o\,d\Omega_o$ over the circular aperture $b\leq10M$.
The gravastar-to-black-hole ratios are $1.0493$ for model I and
$1.2319$ for model II. The enhanced inner illumination therefore
increases the integrated image intensity by approximately $4.93\%$
and $23.19\%$, respectively, for these source prescriptions.

The final radial sampling contains $9610$ rays and resolves up to
the twelfth chronological disk crossing outside the narrow critical
strip. Refining the ray sampling changes the aperture-integrated
intensities by less than $1.1\times10^{-5}$ relative and the pixel
intensities by less than $1.1\times10^{-3}I_*$. Doubling the
radial integration bins changes any displayed pixel by less than
$2.5\times10^{-4}I_*$. The independent crossing and image-integration
checks are included in the accompanying numerical data.
\end{revision}

\section{\rev{Optical degeneracy and higher-order parameter separation}}
\label{sec:optical_identifiability}
\begin{revision}
At fixed $M$, shell radius $a_0$ and interior curvature $H^2$, every
geometric quantity constructed from $F_q$ depends on $(s,e)$ only
through $q$. Two parameter pairs with equal $q$ therefore have the same
shell stresses, matching redshift, photon orbit and ray trajectories
in this approximation. Parameter scans at fixed $s$ or fixed $e$
measure different directions through this one-dimensional family.
We now retain the coefficient $D$ in Eq.~\eqref{eq:id_expansion} to
determine how additional optical information enters.

\subsection{\rev{Photon orbit and logarithmic deflection}}

For $F_4=F_q+D/r^4$, the photon-orbit condition is
\begin{equation}
 r_{\rm ph}^4-3Mr_{\rm ph}^3+2qr_{\rm ph}^2+3D=0.
 \label{eq:id_photon}
\end{equation}
We select the nondegenerate outer maximum of $F_4/r^2$ above the
shell. On the $D=0$ background,
\begin{equation}
 r_0=\frac{3M+\sqrt{9M^2-8q}}2,\qquad
 b_0=\frac{r_0}{\sqrt{F_q(r_0)}}.
 \label{eq:id_background}
\end{equation}
Implicit differentiation of Eq.~\eqref{eq:id_photon} at fixed $q$
gives
\begin{align}
 \left.\frac{\partial r_{\rm ph}}{\partial D}\right|_0
 &=-\frac{3}{r_0^2(2r_0-3M)},\nonumber\\
 \left.\frac{\partial\ln b_c}{\partial D}\right|_0
 &=-\frac1{2r_0^4 F_q(r_0)}.
 \label{eq:id_shifts}
\end{align}
We also define the background logarithmic coefficient
$\bar a_0=(1-2q/r_0^2)^{-1/2}$.
Keeping the exact $q$ background retains its nonlinear contributions
while isolating the first correction in $D$.

For rays that turn outside the photon orbit, the strong-deflection
expansion \cite{Bozza2002NCaudit} is
\begin{equation}
 \Phi_>(b)=-\bar a\ln\left(\frac b{b_c}-1\right)
                  +\bar b_>+o(1),
 \label{eq:id_outer_log}
\end{equation}
with
\begin{align}
 \bar a&=\sqrt{\frac{2}{2F(r_{\rm ph})-r_{\rm ph}^2F''(r_{\rm ph})}}
 \nonumber\\
 &=\left(1-\frac{2q}{r_{\rm ph}^2}-\frac{9D}{r_{\rm ph}^4}\right)^{-1/2}
 \quad\text{for }F=F_4.
 \label{eq:id_abar}
\end{align}
For penetrating rays, the exterior integral crosses both sides of the
potential maximum on each passage. Expanding the radial denominator
quadratically around that maximum yields
\begin{equation}
 \Phi_<(b)=-2\bar a\ln\left(1-\frac b{b_c}\right)
                  +\bar b_<+o(1).
 \label{eq:id_inner}
\end{equation}
This factor of two is the established inner-image result for
ultracompact objects \cite{Shaikh2019NCaudit}. For a shell strictly
inside the nondegenerate photon orbit, the de Sitter contribution in
Eq.~\eqref{eq:Phi_crossing} remains finite and enters $\bar b_<$.
The finite-observer tail $J_{D_o}(b)$ is also smooth at $b_c$, so it
changes the finite strong-deflection terms while preserving these
logarithmic coefficients.
Consequently, consecutive visibility boundaries, separated by $\pi$
in deflection angle, obey
\begin{align}
 \frac{\omega_{n+1}-b_c}{\omega_n-b_c}
       &\longrightarrow e^{-\pi/\bar a},\nonumber\\
 \frac{b_c-\zeta_{n+1}}{b_c-\zeta_n}
       &\longrightarrow e^{-\pi/(2\bar a)}.
 \label{eq:id_spacing}
\end{align}
Thus the exterior and interior boundary sequences give two estimates
of the same local curvature of the photon barrier.

\subsection{\rev{Local separation of the deformation scales}}

On the leading cancellation curve $q=0$, one has $e=16Ms$ and
$D<0$ for every nonzero physical deformation. To first order in
$D/M^4$,
\begin{align}
 r_{\rm ph}&=3M-\frac D{9M^3}+\cdots,\nonumber\\
 b_c&=3\sqrt3M\left(1-\frac D{54M^4}+\cdots\right),\nonumber\\
 \bar a&=1+\frac D{18M^4}+\cdots.
 \label{eq:id_cancel}
\end{align}
The corrections raise $r_{\rm ph}$ and $b_c$ and lower $\bar a$.
The omitted dimensionless contributions are of order $D^2/M^8$
within $F_4$, with an additional contribution of order $E/M^6$ from
the full lapse. The Schwarzschild cancellation at order $r^{-2}$
therefore acquires a calculable residual at order $r^{-4}$.

For an inference setup with independently calibrated mass and
distance, define $Q=q/M^2$ and $\delta=D/M^4$. The linear response
about Schwarzschild is
\begin{equation}
 \begin{aligned}
 \begin{pmatrix}b_c/(3\sqrt3M)-1\\ \bar a-1\end{pmatrix}
 & =\begin{pmatrix}-1/6&-1/54\\1/9&1/18\end{pmatrix}
 \begin{pmatrix}Q\\\delta\end{pmatrix}
 \\
 &\quad+O(Q^2,Q\delta,\delta^2).
 \end{aligned}
 \label{eq:id_jacobian}
\end{equation}
The response matrix has determinant $-7/972$. The map from the
microscopic parameters to the exterior coefficients has determinant
\begin{equation}
 \det\frac{\partial(q,D)}{\partial(s,e)}=-2Me-8\pi^2Ms^2.
 \label{eq:id_micro_jacobian}
\end{equation}
Both maps are locally invertible in the weakly deformed regime with
$M>0$, $s,e\geq0$, away from the undeformed origin. Their composition
shows how $b_c/M$ and $\bar a$ encode independent information about
the two deformation scales. The microscopic inversion becomes
poorly conditioned as $s$ and $e$ approach zero, so a measurement
forecast should propagate mass, distance and image-model uncertainties
through this response matrix.

\begin{table}[t]
\revisioncolor
 \caption{\rev{Two parameter pairs with $q/M^2=0.06$, $a_0/M=2$ and
 $H^2M^2=0.2$. The full exterior is static above each shell and has
 an unstable photon orbit outside it.}}
 \label{tab:id_pairs}
 \centering
 \begin{tabular}{lrr}
 \toprule
 Quantity & Pair A & Pair B\\
 \midrule
 $s/M$ & $0.008$ & $0.012$\\
 $e/M^2$ & $0.008$ & $0.072$\\
 $F(a_0)$ & $0.01499782$ & $0.01495345$\\
 $r_{\rm ph}/M$ & $2.9594560536$ & $2.9595390775$\\
 $b_c/M$ & $5.1435732322$ & $5.1436447942$\\
 $\bar a$ & $1.0069196976$ & $1.0068767948$\\
 \bottomrule
 \end{tabular}
\end{table}

\begin{figure*}[t]
 \centering
 \includegraphics[width=0.95\textwidth]{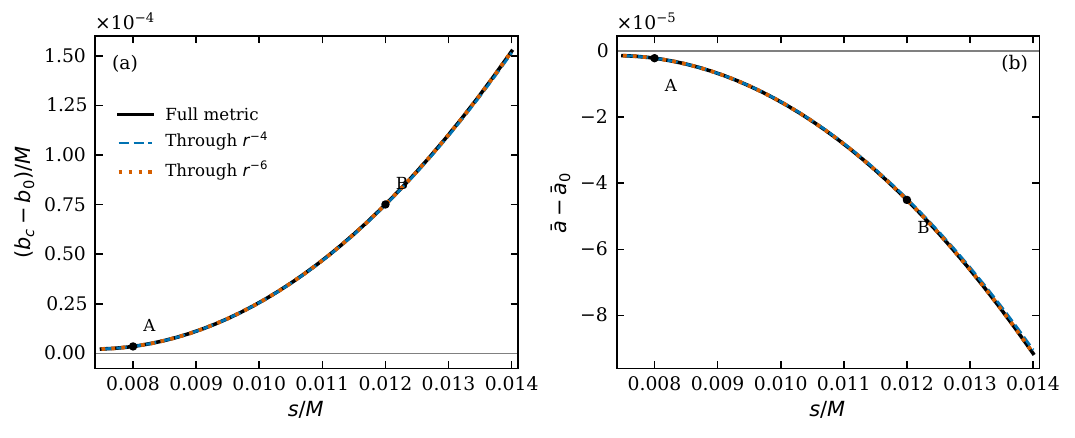}
 \caption{\rev{Optical splitting along the constant-$q$ family
 $q/M^2=0.06$, with $e/M^2=16s/M-0.12$.
 Deviations from the common inverse-square prediction are shown for
 (a) the critical impact parameter and (b) the strong-deflection
 coefficient. The full metric is compared with the expansions through
 $r^{-4}$ and $r^{-6}$. The labeled points are the examples in
 Table~\ref{tab:id_pairs}; the thin horizontal reference represents
 the inverse-square model.}}
 \label{fig:identifiability}
\end{figure*}

Table~\ref{tab:id_pairs} gives two equal-$q$ examples that coincide
exactly in the leading model. With the full lapse their differences
are $\Delta b_c/M=7.15620\times10^{-5}$ and
$\Delta\bar a=-4.29028\times10^{-5}$. The $r^{-4}$ approximation
reproduces these splittings with relative discrepancies of
$0.377\%$ and $0.833\%$, respectively; the $r^{-6}$ term improves
both results. Across the examples, the largest expansion ratios at
the shell are $\pi s/a_0=0.01885$ and $\sqrt e/a_0=0.13416$.
Figure~\ref{fig:identifiability} follows the splitting continuously
along the equal-$q$ family. Direct near-critical integration also
recovers the penetrating-ray slope $2\bar a$ with relative error
below $1.9\times10^{-7}$ for the tabulated pairs.
\end{revision}

\section{\rev{Conclusions}}
\label{Conc}
\begin{revision}
We constructed a thin-shell gravastar from a de Sitter interior and
a phenomenological exterior combining the full Lorentzian mass
profile with the KS factor. Interpreting the exterior through the
Einstein equations supplies a consistent effective source and the
standard junction conditions. The shell is restricted to radii at
which the retained exterior and the interior lapse are positive.

The surface stresses depend on the shell radius, the independent
interior curvature and both deformation scales. For the optical
reference configuration, the weak and strong surface energy
conditions hold, whereas the dominant condition fails. Transparent
ray tracing under rear-plane illumination gives a central illuminated region and alternating
visibility rings. Increasing the KS parameter in the chosen family
moves the exterior photon orbit and critical impact parameter
outward, while the first three interior visibility boundaries move
inward. The higher-order interior boundaries follow the shift of
the critical scale, illustrating the different roles of the shell
redshift and the exterior photon barrier.

The finite-observer construction reproduces the Schwarzschild--de Sitter
reference geodesics and gives a direct comparison with the backlit
black-hole exterior. For the ultracompact reference family, the two
images agree outside the critical circle;
transmission through the gravastar adds the central disk and inner rings.
For the separate fixed-curvature scan with
$\Theta=\hat\eta=0.06$ and $\lambda=0.01$, the exterior photon orbit
is removed from the retained geometry at $a_0/M=2.96068756$.
A dark annulus persists until the first boundary pair merges at
$a_0/M=4.61956378$. Its survival beyond the photon-orbit radius ties
the backlit dark region to the full deflection map and the source
geometry.

For a face-on thin emitting layer, the observed intensity is the sum
of the redshift-weighted contributions from exterior disk crossings.
The gravastar and black hole share the incoming contribution, while
the transparent gravastar adds light from the outgoing trajectory
after core transit. Their intensities agree outside the critical
circle. The two prescribed emission profiles produce different inner
ring brightnesses, and both have a dark central line of sight because
the radial ray misses the exterior emitting layer. These images
connect the core-transmission mechanism to a specified source model.

For the associated horizon-bearing family, the physical horizon
radius determines the ADM mass and surface-gravity temperature.
The effective Einstein description gives area entropy together
with radial-pressure work. At fixed deformation lengths, the
mass-profile identity relates the ADM-mass variation to the
entropy variation. The admissible outer-horizon branches displayed
here have decreasing temperature and negative entropy heat
response, and the exact pure KS limit recovers the Schwarzschild
temperature at fixed mass.

The leading exterior coefficient $q=8M\sqrt{\theta/\pi}-\eta^2/2$
reduces the two microscopic scales to a single geometric parameter.
Retaining the next coefficient $D$ breaks this degeneracy. The
critical impact parameter and logarithmic strong-deflection
coefficient have a nonsingular response to $(q,D)$, and the
full-metric calculation resolves their splitting along an equal-$q$
family. This result identifies the observables and accuracy scales
needed for a parameter-estimation study. Extending the calculation
to a specified shell equation of state, moving emitters and absorption
would connect the equilibrium geometries to dynamical stability and
more detailed emission models.
\end{revision}

\begin{acknowledgments}
We thank CNPq and CAPES for partial financial support. M.A. A.  acknowledges support from CNPq (Grant no. $301683/2025-5$). A. T. N. S. thanks the Paraíba State Research Support Foundation (FAPESQ) (Grant No. $765/2026$) for financial support. L.C. acknowledges support from CNPq (Grant no. $306842/2026-2$). 
\end{acknowledgments}
\appendix
\begin{revision}
\section{\rev{Numerical implementation and reproducibility}}
\label{app:numerics}

The shell and thermodynamic scans evaluate the unexpanded lapse and
its analytic derivatives. Positivity bounds on $\mu-r\mu'$ certify
the retained static exterior and outer-horizon branch used in the
figures. Critical optical quantities follow from bracketed root
finding for the outer photon orbit. The ray integrals are evaluated
with endpoint-regularizing substitutions and adaptive quadrature;
the independent high-precision checks are reported
in Sec.~\ref{sec:null_geodesics}. The observer endpoint is retained
explicitly through the exterior tail integral. The reference
benchmark uses two finite endpoints, while the NC--KS source-plane
images use one finite observer and an asymptotic source. The radius
scan follows the change in the entry threshold at $a_0=r_{\rm ph}$
and evaluates the grazing limit at each ring merger.
The equal-$q$ calculations use
45-digit arithmetic. Differencing the penetrating phases at
$1-b/b_c=10^{-7}$ and $10^{-8}$ measures the coefficient in
Eq.~\eqref{eq:id_inner}. Binary images are constructed from the
computed visibility boundaries at the resolution given in
Fig.~\ref{fig:optical_visibility}. The shell-radius scan samples
124 admissible radii with positive surface density. Root tables,
endpoint comparisons and numerical convergence checks are supplied
with the figures.

The disk calculation evaluates the exterior equatorial crossings of
backward rays from the polar observer and adds the local bolometric
emission with the finite-observer redshift factor. Interior transit
uses the analytic de Sitter angle. Captured black-hole trajectories
retain their disk intersections before terminating at the horizon.
The intensity sampling resolves the distinct transfer branches and
the jumps at the prescribed emission edges. Convergence tests compare
ray integration tolerances and image resolution near the critical
circle, as detailed in Sec.~\ref{sec:disk_images}. Pixel integration
uses analytic intersections of radial annuli with square image pixels.
In the strip $|1-b/b_c|<10^{-7}$, we match the finite part of the
intensity to the directly integrated boundary values and continue
the local logarithmic term. For model II its coefficient is
$\bar a I_*[F_+(r_{\rm ph})/F_+(D_o)]^2/\pi$ for both objects
above $b_c$ and for the gravastar below it. The captured black-hole
branch below $b_c$ has half this coefficient because its emitting
segment covers the outer half of the photon-orbit bottleneck.
The coefficient vanishes for model I, whose inner edge is outside
the photon orbit. Halving the strip and adding direct rays changes
the integrated intensities by less than $2.9\times10^{-7}$ relative
and any tested pixel by less than $2.1\times10^{-5}I_*$.

OpenAI Codex, using a GPT-6-based model, assisted with symbolic and
numerical code development, calculation checks, literature verification,
and manuscript revision, including the higher-order optical analysis
and the disk transfer-function calculation.

\end{revision}

\end{document}